\documentclass[twocolumn,prx,amsmath,amssymb,superscriptaddress,longbibliography,floatfix]{revtex4-2}

\usepackage[T1]{fontenc}
\usepackage{bbold}

\usepackage{mathtools}
\usepackage{physics, calc} 
\usepackage{graphicx}
\usepackage{tabularx}
\usepackage{colortbl}
\usepackage[citecolor=blue, urlcolor=blue, linkcolor=blue, colorlinks=true]{hyperref}

\usepackage{txfontsb}
\usepackage{times}
\usepackage{mathrsfs}
\usepackage{dsfont}

\graphicspath{{figures/}}

\newcommand*\dagg{^{\dagger}}	
\newcommand{\C}{\mathbb{C}}

\newcommand*{\todo}[1]{\textcolor{red}{#1}}

\RequirePackage[capitalize]{cleveref}
\crefformat{section}{Sec.~#2#1#3}

\begin{document}
\title{Topological two-dimensional Floquet lattice on a single superconducting qubit}
\author{Daniel Malz}
\thanks{Both authors contributed equally.}
\affiliation{Max Planck Institute for Quantum Optics, Hans-Kopfermann-Straße 1, D-85748 Garching, Germany}
\affiliation{Munich Center for Quantum Science and Technology, Schellingstraße 4, D-80799 München, Germany}
\author{Adam Smith}
\thanks{Both authors contributed equally.}
\affiliation{Department of Physics, TFK, Technische Universit{\"a}t M{\"u}nchen, James-Franck-Stra{\ss}e 1, D-85748 Garching, Germany}
\affiliation{School of Physics and Astronomy, University of Nottingham, Nottingham, NG7 2RD, UK}
\affiliation{Centre for the Mathematics and Theoretical Physics of Quantum Non-Equilibrium Systems, University of Nottingham, Nottingham, NG7 2RD, UK}

\begin{abstract}
  Current noisy intermediate-scale quantum (NISQ) devices constitute powerful platforms for analogue quantum simulation. 
  The exquisite level of control offered by state-of-the-art quantum computers make them especially promising to implement time-dependent Hamiltonians.
  We implement quasiperiodic driving of a single qubit in the IBM Quantum Experience and thus experimentally realize a temporal version of the half-Bernevig-Hughes-Zhang Chern insulator.
  Using simple error mitigation, we achieve consistently high fidelities of around 97\%.
  From our data we can infer the presence of a topological transition, thus realizing an earlier proposal of topological frequency conversion by Martin, Refael, and Halperin.
  Motivated by these results, we theoretically study the many-qubit case, and show that one can implement a wide class of Floquet Hamiltonians, or time-dependent Hamiltonians in general.
  Our study highlights promises and limitations when studying many-body systems through multi-frequency driving of quantum computers.
\end{abstract}

\maketitle

\emph{Introduction.---}Noisy intermediate-scale quantum (NISQ) computers may not yet offer fully fault-tolerant quantum computing facilities, but they nevertheless constitute a versatile experimental platform with the potential for fundamental research, small-scale computation or quantum simulation~\cite{Preskill2018}.
The typical model of a quantum computer is that of a quantum circuit, which is a sequence of gates applied to the qubits~\cite{Nielsen2010}.
In principle, the time-evolution of any many-body quantum systems can be simulated by applying a Trotterization, which turns continuous time evolution into a discrete local quantum circuit~\cite{Lloyd1996}. This results in a digital quantum simulation, which has been benchmarked for a range of different models on existing quantum computers~\cite{Smith2019a,Tacchino2019,Chiesa2019,Arute2020}.

In superconducting circuits, the currently leading technology, quantum circuits are constructed from a set of available gates, which correspond to a set of carefully calibrated microwave pulses applied to its input ports~\cite{Krantz2019}.
The abstraction into quantum circuits hides the complexity of the underlying many-body system, whose continuous evolution
offers exciting directions in \emph{analogue} quantum simulation~\cite{Cirac2012,Schmidt2013}, which potentially incurs significantly less overhead.
This perspective has been explored in a series of theoretical and experimental works~\cite{Roushan2016,Roushan2017,Chiaro2019,Guo2019a,Yanay2020}.
If the intrinsic many-body nature of quantum computers is combined with the capacity to apply essentially arbitrary drives, they may serve also as powerful analogue quantum simulators for very large classes of time-dependent Hamiltonians.

The evolution under time-dependent Hamiltonians is incredibly rich and exhibits many novel phenomena, even at the level of individual qubits.
A particular example is the temporal topological transition that occurs in the presence of quasiperiodic driving, theoretically predicted by Martin, Refael, and Halperin in 2017~\cite{Martin2017}. Using a Floquet treatment of the driven qubit, the dynamics is related to the properties of a two-dimensional lattice model, the half-Bernevig-Hughes-Zhang (BHZ) Chern insulator~\cite{Bernevig2006}. As a function of time, the driven qubit explores the whole Brillouin zone, which causes the work done by the two drives to be quantized and proportional to the integer Chern number, which is determined by the parameters in the drives.
Recently, temporal topology has been classified, analogously to the classification of topological insulators~\cite{Crowley2019}, and extension to larger systems, such as spin--resonator systems~\cite{Nathan2020}, or two-qubit systems with interactions~\cite{Korber2020} have been proposed.
While quasiperiodic driving typically maps to systems without boundary, one can in principle also introduce boundaries through quantum feedback~\cite{Baum2018}.

In this work we experimentally demonstrate this temporal topological behaviour of a single qubit on an existing quantum device, using continuous driving, implemented with the fine-grained access offered by QISKIT pulse~\cite{Alexander2020}.
Choosing a specific driving with two incommensurate frequencies, we observe a topological transition in the temporal dynamics of the qubit, finding good agreement with simulations.
Despite achieving high fidelities of around 97\% after error mitigation, the Chern number inferred from the measured frequency conversion shows much larger errors. We develop a simple noise model that explains and reproduces this effect.

Motivated by this experiment, we theoretically derive the class of Hamiltonians that can readily be implemented on state-of-the-art quantum computers.
As one concrete example, this offers an exciting perspective to study strongly interacting Floquet systems~\cite{Eckardt2017,Oka2019,Ozawa2019a,Ozawa2019,Khemani2019} with an exquisite level of control.
Site-selective control as well as high-fidelity single-site readout confers quantum computers certain advantage over other quantum simulators based on light~\cite{Noh2017} or ultracold atoms~\cite{Bloch2012}, making them ideally suited for the analogue quantum simulation of generic many-body time-dependent Hamiltonians.

\emph{Theoretical description.---}In quantum computers based on superconducting transmon-like qubits~\cite{Koch2007,Kjaergaard2019}, the Hamiltonian describing a single qubit can be cast in the form of a Duffing oscillator with driving
\begin{equation}
  H(t) = \omega_0a\dagg a+Ua\dagg a\dagg aa+(a+a\dagg)D(t).
  \label{eq:duffing}
\end{equation}
The qubit frequency is denoted $\omega_0$ and $U$ quantifies the anharmonicity that separates the lowest two levels that define the qubit from the higher levels of the superconducting circuit.
The ideal drive signal is parameterized as~\cite{Alexander2020}
\begin{equation}
  \begin{aligned}
  	D(t) &= \frac{\Omega_{\mathrm{max}}}{2}\Re\left[ e^{i(\omega_0+\Delta_c) t}d(t) \right],
  \end{aligned}
  \label{eq:drive_parameterization}
\end{equation}
where $\Omega_{\mathrm{max}}$ denotes the maximum Rabi frequency attainable in the system, and $\omega_0+\Delta_c$ is the carrier frequency. In the following, we choose the detuning $\Delta_c=0$.
Anticipating our later interpretation as a spin-$1/2$ particle in a time-dependent field, we parameterize the dimensionless drive shape $d(t)\in\C, |d(t)|\leq1$ in terms of the dimensionless magnetic field $\tilde h_+(t) = \tilde h_x(t)+i\tilde h_y(t)$
\begin{equation}
  d(t) \equiv \tilde h_+(t) \exp i\phi(t),\quad
  \phi(t) = -2\Omega_{\mathrm{max}}\int_0^t\tilde h_z(t')dt'.
  \label{eq:d}
\end{equation}

To treat the system as a qubit, the maximum drive strength needs to be much weaker than the anharmonicity $\Omega_{\mathrm{max}}\ll U$.
Assuming this is fulfilled, applying a rotating-wave approximation and passing into a frame rotating with respect to the time-dependent Hamiltonian $H_z(t)=(\omega_0/2-h_z(t))\sigma_z$, we rewrite \cref{eq:duffing} as~\footnote{See Appendix for additional derivations, simulation and experimental details, as well as more data.}
\begin{equation}
  H_{\mathrm{spin}}(t) = \vec h(t) \cdot\vec \sigma.
  \label{eq:H_spin}
\end{equation}
Note the subtlety in the chosen rotating frame.
Instead of changing the qubit frequency in the lab frame, we re-parameterize time by passing into a rotating frame with respect to $H_z(t)$.
This constitutes a continuous version of virtual $Z$ gates~\cite{McKay2017}.

\emph{Topological frequency conversion.---}Given \cref{eq:H_spin}, we can now realize the proposal by Martin \emph{et al.}~\cite{Martin2017}.
We apply the quasi-periodic time-dependent magnetic field
\begin{equation} 
\begin{aligned}
  H(t) = \eta \Big\{& \sin(\omega_1 t + \phi_1) \sigma_x + \sin(\omega_2 t + \phi_2) \sigma_y \\
  &+ \left[M - \cos(\omega_1t + \phi_1) - \cos(\omega_2t + \phi_2)\right]\sigma_z  \Big\}
\end{aligned}
\label{eq:h(t)}
\end{equation}
where the ratio $\omega_2/ \omega_1$ should be irrational.
In the following, we set $\omega_2/\omega_1=(1+\sqrt{5})/2$.
We consider the model in the strong drive limit, i.e., $\eta \gg \omega_1,\omega_2$. In this limit, a Floquet ansatz reveals a direct connection to the two-dimensional half-BHZ Chern insulator~\cite{Bernevig2006} with a constant electric field applied~\cite{Martin2017,Note1}.
In the strong-drive limit, the electric field is weak and leads to a slow adiabatic evolution of the initial wavepacket through the Brillouin zone. 
During this evolution, which explores the whole Brillouin zone, the system effectively measures the Chern number, which results in a topologically quantized energy pumping rate from one drive to the other~\cite{Martin2017}
\begin{equation}\label{eq:work quantization}
    \pi\frac{(W_1 - W_2)}{\omega_1 \omega_2 T} = C,
\end{equation}
where $W_i$ is the work done by the $i$th drive, defined below [\cref{eq:W_i}].
As a function of $M$, the system undergoes a topological transition in which the Chern number changes.

To determine the pumping rate experimentally, we measure the work done by each of the drives. If we first split the Hamiltonian into the two contributions from each drive,
\begin{equation}
    H(t) = h_1(t) + h_2(t) + \eta M \sigma_z,
\end{equation}
then the work done by each drive over a period $T$ is given by
\begin{equation}
  W_i(T) = \int^T_0 dt\, \langle \Psi(t) | \frac{\text{d} h_i (t)}{\text{d} t} |\Psi (t) \rangle,
  \label{eq:W_i}
\end{equation}
where $|\Psi(t)\rangle$ is the state of the qubit at time $t$ evolving under the time-dependent Schr{\"o}dinger equation $i\partial_t |\Psi(t)\rangle = H(t) |\Psi(t)\rangle$, where $|\Psi(t=0)\rangle$ is chosen to be an instantaneous eigenstate of $H(t=0)$.

\emph{Experiment.---}Our experimental protocol consists of four main pulse sequences. First, we initialize the qubit in the instantaneous eigenstate of Hamiltonian Eq.~\eqref{eq:h(t)} at $t=0$. This is achieved using an IBM calibrated pulse sequence to perform a general qubit rotation. Second, the main pulse sequence implements the time-dependent Hamiltonian~\eqref{eq:h(t)}, with the drive parameterized as in Eq.~\eqref{eq:drive_parameterization}.
We perform experiments with different drive lengths of up to 20$\mu$s to obtain 800 data points.
Third, we apply an IBM calibrated pulse to change the basis. For each drive length, we rotate into the X, Y, and Z bases in order to perform full state tomography. Finally, we perform single-shot projective measurements of the qubit using an IBM pre-calibrated readout pulse sequence. We average over 8192 shots for each observable corresponding to a statistical error of approximately $1\%$.

We fix $\omega_2 = \varphi \omega_1$, where $\varphi = (1+\sqrt{5})/2$ is the golden ratio.
Ideally, we would be like to set the ratio $\omega_1/\eta$ as small as possible, to get as close as possible to adiabatic evolution.
However, due to the finite coherence time $\tau\gtrsim 100\mu$s of the IBM Q device, we must choose $\omega_1^{-1} \ll \tau$.
In terms of the maximum Rabi frequency $\Omega_{\mathrm{max}}$, we choose $\eta = 0.9 \Omega_\text{max}$, in order to avoid driving transitions to higher excited states. Using numerical simulations we found the best compromise was to choose a total simulation time of $20\mu$s and set $\omega_1 = 0.125 \eta$, which corresponds to $\omega_1^{-1} \approx 240$ns.
In order to improve the fidelity of our results we start the drive with a linear ramp of the drive frequencies $\omega_1, \omega_2$ over a period of 444ns. This reduces transient effects and reduces high-frequency Rabi oscillations seen in the simulation results. The single-qubit IBM device we used, codenamed armonk, had qubit frequency $\omega_0 \approx 4.97$ GHz and maximum Rabi frequency of $\Omega_\text{max} \approx 36.9$ MHz.

\begin{figure}[tb]
 \includegraphics[width=\columnwidth]{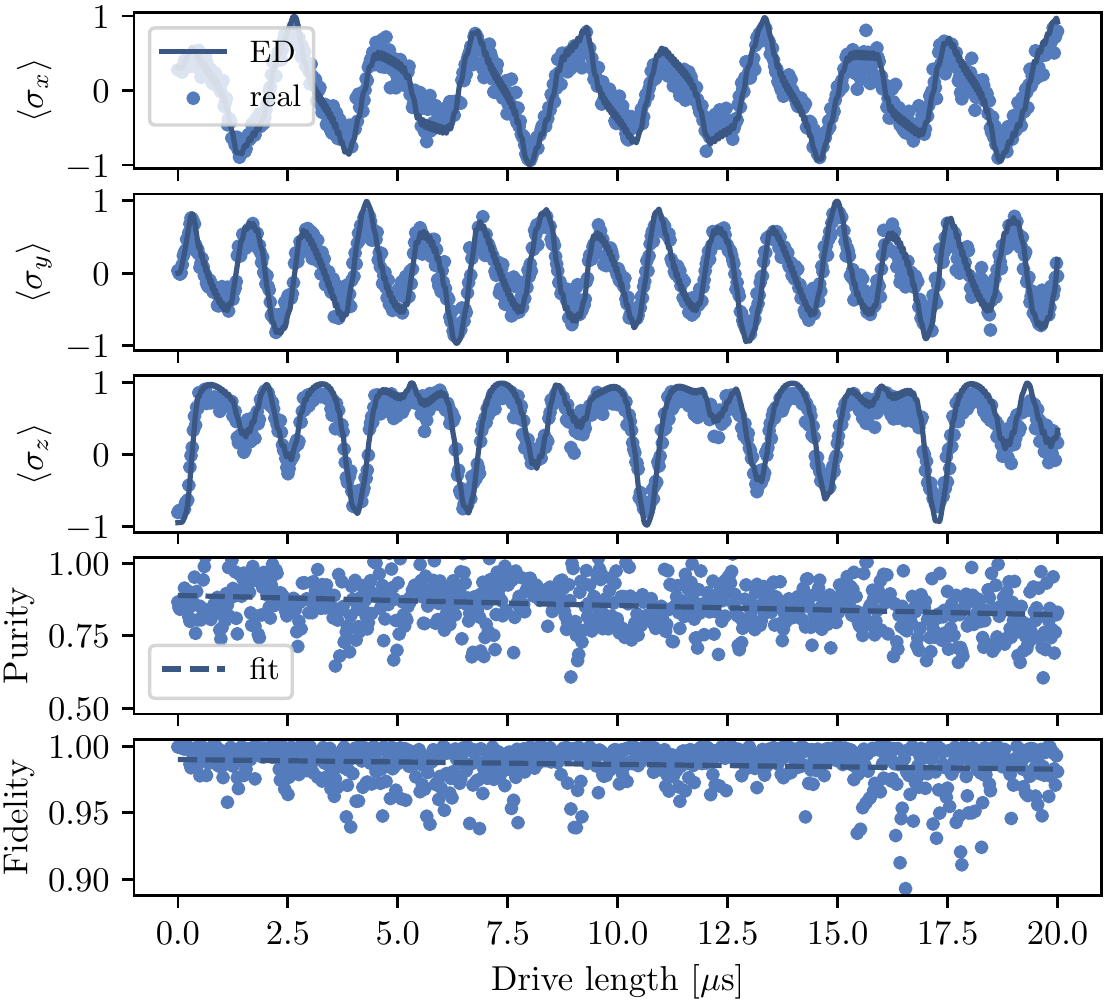}
\caption{
  Tomography data for $M=1$, $\omega_1 = 0.125$.
  The top three panels compare experimentally measured Pauli expectation values (projected back onto the Bloch sphere) against exact numerical simulation.
  We also show the purity of the measured state, which we fit by the function $1/2 + ae^{-t/\lambda}$, where $a=0.387$ and $\lambda \approx 109 \mu$s is compatible with the device coherence time as measured by IBM.
  Note that $a<0.5$ as the measurement sequence take a finite time, such that the qubit has lost purity even when the simulation time is zero.
  The bottom panel shows the fidelity $\mathcal{F} = |\langle\psi|\phi\rangle|^2$ between the measured state projected onto the Bloch sphere, $|\psi\rangle$, and the numerically simulated state, $|\phi\rangle$. We fit the fidelity with $a e^{-t/\xi}$, where $a\approx 0.99$ and $\xi\approx2.71 ms$, which verifies the effectiveness of error mitigation by projecting onto the Bloch sphere.
  A significant portion of the lost fidelity after error mitigation can therefore be attributed to statistical error, which is 2\% in our case.
}
\label{fig:tomography}
\end{figure}

With the above experimental protocol we measure the observables in the frame rotating with the qubit frequency. Since the Hamiltonian \cref{eq:H_spin} is only realized in a given time-dependent reference frame, we must additionally perform a virtual-Z rotation, which we achieve by post-processing the data to apply the rotation
\begin{equation}
\begin{aligned}\label{eq: z rotation}
  \langle \sigma_x\rangle_\text{rotating} &= \cos \phi(t) \langle \sigma_x\rangle + \sin \phi(t) \langle \sigma_y \rangle, \\
  \langle \sigma_y\rangle_\text{rotating} &= -\sin \phi(t) \langle \sigma_x\rangle + \cos \phi(t) \langle \sigma_y \rangle,
\end{aligned}
\end{equation}
where $\phi(t)$ is given in Eq.~\eqref{eq:d}. When post-processing the data, we additionally mitigate some of the errors due to decoherence and measurement error by projecting the measured qubit density matrix onto the Bloch sphere. We find that this method of error mitigation improves quantitative agreement with numerical simulations in all cases we considered. This effectively removes the effects of the dominant depolarizing channel~\cite{Vovrosh2021}.
With this error mitigation strategy, the average fidelity across all experiments is 0.971 (excluding $M=1.7, 2.3$ due to strong diabatic effects that arise due to the proximity to the gap closing at $M=2$~\cite{Note1}).

The experimental tomography data is shown in Fig.~\ref{fig:tomography} for $M=1$, which closely matches the exact numerical simulation. From this data we can also compute the purity of the measured state $\text{Tr}(\rho^2)$, also shown in Fig.~\ref{fig:tomography}.
The measured purity is less than one due to decoherence over the course of the experiment along with the significant $\sim 3\%$ measurement error. Fitting the purity with an exponential decay, we extract a decoherence time of $\lambda\approx 109 \mu$s (at $M=1$), which is consistent with the $T_1, T_2 \gtrsim 100 \mu$s decoherence times measured by IBM.

\begin{figure}[tb]
 \includegraphics[width=\columnwidth]{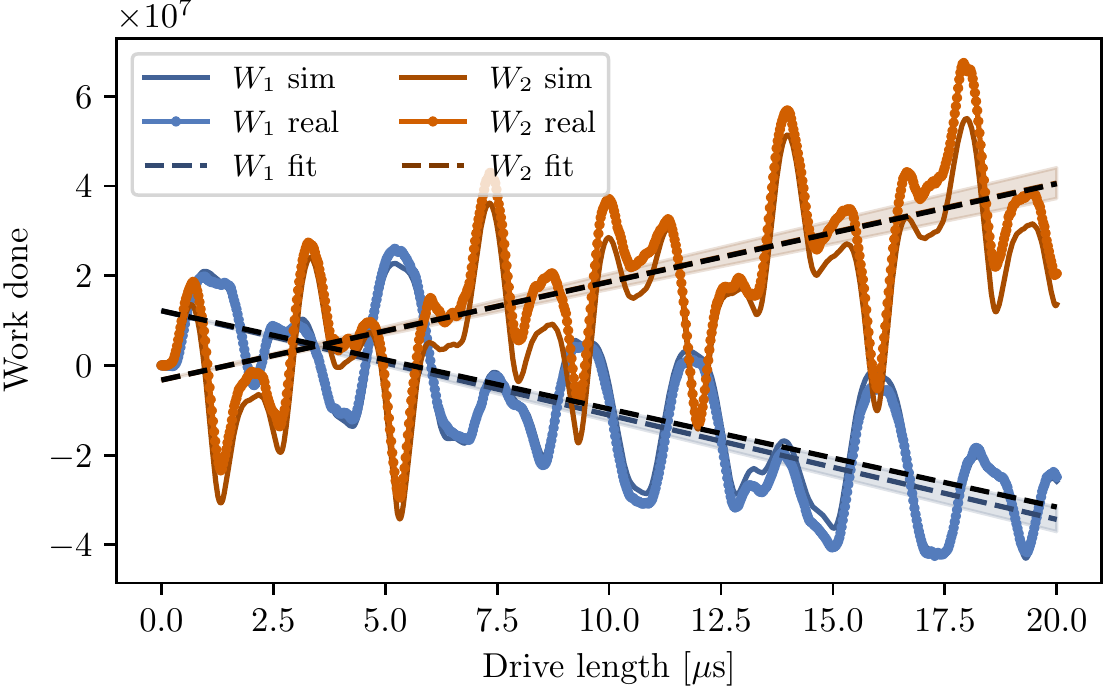}
\caption{Work done by the two incommensurate drives, calculated using \cref{eq:W_i}. The experimental data for $M=1$, $\omega_1=0.125$ measured at 800 points is compared with a numerical simulation of the same setup. The experimental data is fitted with a line (coloured dashed lines) using least-squares regression, for which the 95\% confidence interval for the slope is shown as the coloured region. The expected slope is shown as black dashed lines.}
\label{fig:frequency_conversion}
\end{figure}

\emph{Results.---}To obtain the work done by each drive [\cref{eq:W_i}], we first compute $\langle \text{d}h_i(t) / \text{d} t \rangle = \text{d}\vec{h}_i(t) / \text{d} t \cdot \langle \vec{\sigma} \rangle$ from the data.
We then perform the integration in Eq.~\eqref{eq:W_i} numerically. \Cref{fig:frequency_conversion} shows the experimental results for the work done, for the case of $M=1$.
This corresponds to a phase of the BHZ model with Chern number $C=-1$, resulting in an average linear decrease (increase) of $W_1$ ($W_2$). The experimental results are in good quantitative agreement with numerical simulations.
Furthermore, by fitting a linear curve to this data using least-squares regression, the slope is in close agreement to that predicted by \cref{eq:work quantization}.

\begin{figure}[tb]
  \includegraphics[width=\columnwidth]{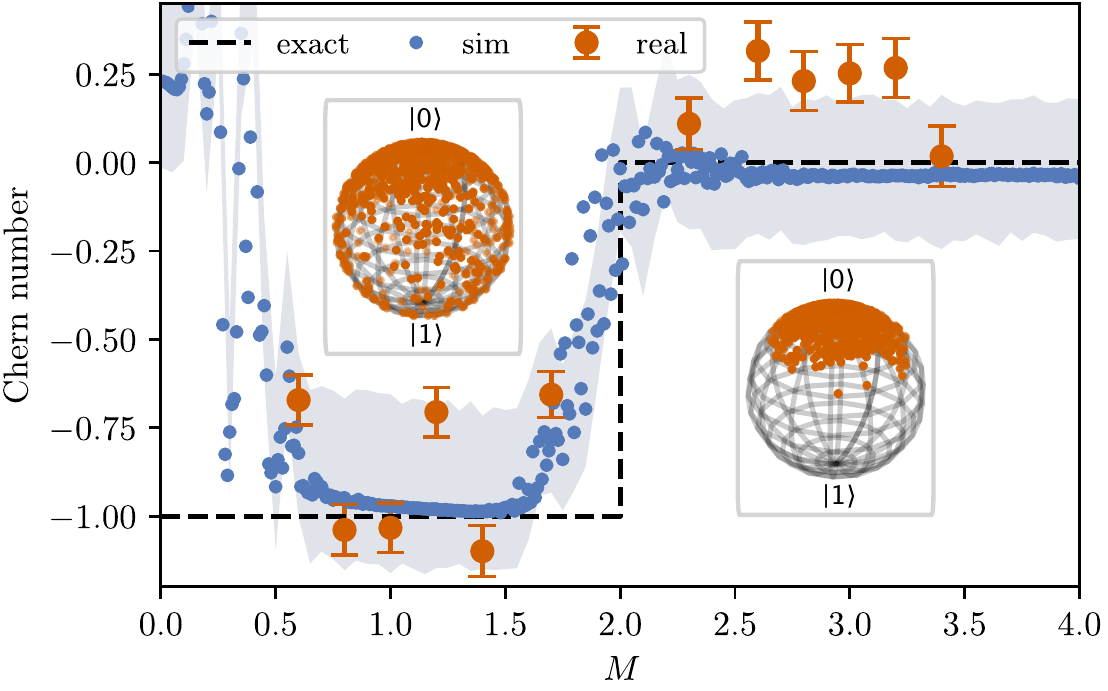}
\caption{
  Experimentally extracted Chern number as a function of $M$. The Chern number is extracted from the linear least-squares regression fit to the work done by each drive, and using \cref{eq:work quantization}. The error bars correspond to the 95\% confidence intervals for the fit. The experimental data is compared against the numerical simulation of the same setup (blue dots) and with the exact value of the Chern number for the corresponding phase of the BHZ model (dashed black curve).
  The shaded blue region corresponds to a simple heuristic error model based on the average fidelity 0.971 of the error mitigated experimental data~\cite{Note1}.
  This error model confirms that the comparatively large deviations of the measured Chern number that we observe are produced already from low loss in fidelity ($\approx3\%$).
  Insets show the distribution of experimental data on the Bloch sphere under dynamics for $M=1$ and $M=3$, illustrating qualitative differences of the time evolution in each of the two phases $|M|<2$ and $|M|>2$.
}
\label{fig:chern_number}
\end{figure}

\Cref{fig:chern_number} shows the extracted Chern number for a range of values of $M$, compared against numerical simulations of the same setup as well as the ideal result in the strong-drive limit. We find reasonable quantitative agreement with the simulations. Furthermore, the transition between different phases with different Chern number is clearly visible in the experimental data. Beyond the extracted value of the Chern number, the qualitative difference between the phases with $|M| < 2$ and $|M| > 2$ is clearly seen by considering the portion of the Bloch sphere covered under time evolution, shown inset \cref{fig:chern_number}. For $|M| < 2$ we observe that the state of the qubit explores the full Bloch sphere under the dynamics of \cref{eq:h(t)} reflecting that the surface traced by the time dependent state has non-trivial winding around the origin and hence non-zero Chern number. When $|M| > 2$ the state of the qubit is instead restricted to either the north or south hemisphere of Bloch sphere and does not wind around the origin and the Chern number is zero.

While overall the agreement is good, the error in the measured Chern number exceeds the error predicted by the fidelity considerably.
We find that this can be explained by a simple error model, in which we take the state predicted in the ideal scenario and randomly perturb its direction on the Bloch sphere to reproduce the measured distribution of fidelities with average $0.971$~\cite{Note1}.
Note that the perturbations at data points corresponding to different times are independent, as they correspond to different experimental runs. 
This perturbation translates into an error in the measured Chern number, whose standard deviation we plot as a shaded region in \cref{fig:chern_number}.
This shows that a error rate consistent with experimental data explains the large errors in the extracted Chern number.

\emph{Extension to qubit arrays.---}In view of the vast research effort in Floquet matter, the question is pertinent whether the presented scheme generalizes to several qubits. To answer this question we must be aware that the coupling between two qubits can be engineered in many ways, as one can choose between direct capacitive or inductive coupling, or indirect coupling via a resonator~\cite{Blais2004,Liu2006,Niskanen2007,Allman2010,Srinivasan2011}. Indeed, alternative approaches have been taken by IBM (see, e.g., Ref.~\cite{Kandala2019}) and Google~\cite{Chen2014c}, with each implementation having their own (dis-)advantages.

We consider two cases in particular. 
The first is when both qubit frequencies and their coupling is fixed, which is relevant for the devices developed by IBM. In this case, the coupling has to be engineered with time-dependent driving of the qubits in order to implement resonant processes to second order in the Hamiltonian~\cite{Paraoanu2006,Rigetti2010,DeGroot2010}.
This technique produces tuneable $XZ, YZ, ZX$, and $ZY$ interactions, which are used to engineer CR gates~\cite{Rigetti2010,Sheldon2016}.
Implementing a similar single-qubit driving as before, and passing into a time-dependent reference frame, we derive an effective Hamiltonian of the form~\cite{Note1}
\begin{equation}
  H_{\mathrm{int}}^{(2)} = \sum_{\langle ij\rangle }g_{ij}(t)\sigma_z^{(i)}\sigma_+^{(j)}+\mathrm{H.c.}
  \label{eq:array_ham1}
\end{equation}
On bipartite lattices, a (virtual) rotation of every second spin maps this to either an Ising interaction or (in general anisotropic) XY interactions, making this technique very versatile.
A drawback is that the Hamiltonian \eqref{eq:array_ham1} is obtained to second order in the original Hamiltonian and by neglecting quickly rotating terms.
It is thus an approximation and care has to be taken that all the steps in its derivation are valid. 
Nevertheless, these conditions can usually be fulfilled through careful choice of the driving parameters.
It may further be possible to actively counteract unwanted effects from this approximation, analogous to the DRAG scheme used to improve the fidelity of digital quantum gates~\cite{Krantz2019}.

If tuneable interactions are available and the qubits can be brought into resonance~\cite{Chen2014c}, the driving scheme is simplified and one readily arrives at the general spin Hamiltonian
\begin{equation}
  H_{\mathrm{array}}(t) = \sum_i\vec h^{(i)}(t) \cdot\vec \sigma^{(i)}
  +\sum_{\langle ij\rangle }g_{ij}(t)\left( \sigma_x^{(i)}\sigma_x^{(j)}+\sigma_y^{(i)}\sigma_y^{(j)} \right)
  \label{eq:array_ham2}
\end{equation}
with a magnetic field along $z$ that takes the form
\begin{equation}
  h_z^{(i)}(t)=h_z(t)+\omega_i(t)-\omega_0,
  \label{eq:mag_field_z}
\end{equation}
where $\omega_i$ are the qubit frequencies and $\omega_0$ is some arbitrarily chosen reference frequency.
We note that tuneable qubit frequencies feature in many implementations~\cite{Dicarlo2009,Barends2014,Kelly2015,Reagor2018}, but not in all~\cite{McKay2016}.
In the latter case---of fixed qubit frequencies but tuneable interactions---one option might be to drive each qubit with a far off-resonant drive to induce an drive-strength-dependent AC Stark shift.

When taking into account the second excited state of each qubit, the qubit array Hamiltonian~\eqref{eq:array_ham2} maps to a Bose-Hubbard model with time-dependent hopping and freely tuneable site-dependent drive and disorder.
We note that the interpretation as a (time-independent) Bose-Hubbard model has enabled the experimental measurement of microscopic features of the many-body localized phase~\cite{Chiaro2019}.
Moving to periodically varying hoppings could allow one to study many-body Floquet models and implement quench and ramp experiments from carefully prepared initial states.

\emph{Discussion.---}While our experiment showcases some of the promises of analogue quantum simulation with time-dependent Hamiltonians, it also highlights some of the difficulties that need to be overcome on the way. 
Concretely, the experimental results shown in \cref{fig:chern_number} would improve with greater coherence time of the qubit, as this would allow us to reduce the modulation frequencies $\omega_1$ and $\omega_2$ in the Hamiltonian, which in turn would improve the strong-driving and adiabatic-modulation approximations.
This is particularly important close to the topological transition at $M=2$ where the gap closes and adiabatic evolution thus requires increasingly long time scales.

In this specific experiment, we also encountered the problem that the measured experimental signature is sensitive to even small amounts of noise.
Recently, Boyers \emph{et al.}~\cite{Boyers2020} demonstrated experimental measurement of the Chern number in the same model using a nitrogen vacancy centre. They determined the Chern number by measuring the Berry curvature directly, whereas here we extract the Chern number from the topological pumping as originally proposed. As a result, they did not observe the substantial errors in the extracted Chern number that we found.
Nevertheless, as we have demonstrated, the temporal topological nature of the qubit time evolution can clearly be extracted despite experimental shortcomings.
With the current progress in superconducting-qubit technology, we expect that our understanding and the fidelity of available quantum computers to increase rapidly, allowing for more complex experiments, in particular with more qubits.

Going beyond (quasi)periodic driving, the capability to engineer time-dependent many-body Hamiltonians offers many exciting perspectives to investigate non-equilibrium physics. 
For example, ramping through a quantum phase transition might allow one to study Kibble-Zurek scaling or in general the dynamics of phase transitions such as the superfluid to Mott insulator transition~\cite{Braun2015}.
Slow variation could also be used to explore adiabatic algorithms and departures from them.
Many-body non-equilibrium physics is notoriously difficult to study with classical computers and this is therefore a prime area of applicability for quantum simulators. 
Noisy intermediate-scale quantum computers offer a versatile combination of single-site control and readout, and large enough system sizes, and thus promise to support and complement analytical and computational approaches to understand many-body non-equilibrium physics. As we have demonstrated with the single-qubit experiment, this is a realistic outlook.

\begin{acknowledgments}
DM acknowledges funding from ERC Advanced Grant QENOCOBA under the EU Horizon 2020 program (Grant Agreement No. 742102). AS was supported by the European Research Council (ERC) under the European Union's Horizon 2020 research and innovation program (grant agreement No. 771537), and in the latter stages of this work by a Research Fellowship from the Royal Commission for the Exhibition of 1851. The views expressed are those of the authors and do not reflect the official policy or position of IBM or the IBM Quantum Experience team.
\end{acknowledgments}

\vspace{10pt}
\begin{center}
    \textbf{\small DATA AVAILABILITY}
\end{center}

All of the data presented in this paper, along with the python code used to generate this data and the figures in this paper, can be found in a public repository~\footnote{Daniel Malz, and Adam Smith, \url{https://doi.org/10.5281/zenodo.4560106}}.

\appendix

\section{Qubit array Hamiltonian}\label{app:array_extension}

\subsection{Details on the single qubit}\label{app:single_qubit_details}

In order to treat the system as a qubit, the maximum drive strength needs to be much weaker than the anharmonicity $\Omega_{\mathrm{max}}\ll U$.
Assuming this is fulfilled, applying a rotating-wave approximation and passing into a frame rotating with respect to the time-dependent Hamiltonian $H_z(t)=(\omega_0/2-h_z(t))\sigma_z$, we rewrite \cref{eq:duffing} as
\begin{equation}
  \begin{aligned}
  	H_{\mathrm{qubit}}(t) &= \frac{1}{2}\sigma_-\left( h_x(t)+ih_y(t) \right)e^{-2i\int_0^t h_z(t')dt'} + \mathrm{H.c.},
  \end{aligned}
  \label{eq:h(t)_lab}
\end{equation}
where we have now introduced the dimensionful magnetic field $\vec h(t) = \Omega_{\mathrm{max}}\vec{\tilde h}(t)$ and the spin lowering operator $\sigma_-$.
If we pass (again) into a rotating frame with respect to the time-dependent Hamiltonian $H_z(t) = -h_z(t)\sigma_z$, the Hamiltonian becomes that of a spin-1/2 particle in a time-dependent magnetic field
\begin{equation}
  H_{\mathrm{spin}}(t) = \vec h(t) \cdot\vec \sigma.
  \label{eq:H_spin_app}
\end{equation}
Note the subtlety when passing from \cref{eq:h(t)_lab} to \cref{eq:H_spin}.
Instead of changing the qubit frequency in the lab frame, we re-parameterize time by passing into a rotating frame with respect to $H_z(t)$.
This constitutes a continuous version of virtual $Z$ gates~\cite{McKay2017}.

\subsection{Fixed qubit frequencies and fixed coupling}

If neither the qubit frequencies nor the coupling between them is tuneable, the interaction Hamiltonian is time-independent. Restricting directly to the qubit subspace and assuming capacitive coupling for simplicity, we have
\begin{equation}
  H(t) = \sum_i\omega_i\sigma_z^{(i)} + D_i(t)\sigma_x^{(i)} + \sum_{\langle ij\rangle }J_{ij}\sigma_x^{(i)}\sigma_x^{(j)}.
  \label{eq:fixed_H}
\end{equation}
If the coupling $J_{ij}$ is weak compared to the relative detuning, the qubits are effectively decoupled.
Coupling between the $i^{\mathrm{th}}$ and $j^{\mathrm{th}}$ qubit can be turned on by driving either one at the frequency of the detuning between them~\cite{Paraoanu2006,Rigetti2010}.

To see this, let us pass to the frame rotating with the qubit frequencies, and neglect terms rotating at the qubit frequency, which yields
\begin{equation}
  H_{\mathrm{RWA}}(t) = \sum_i d_i(t)\sigma_+^{(i)} + 
  \sum_{\langle ij\rangle }J_{ij}\sigma_+^{(i)}\sigma_-^{(j)}e^{i(\omega_i-\omega_j)t}+\mathrm{H.c.},
  \label{eq:fixed_RWA}
\end{equation}
where we parameterize the drive in two steps 
\begin{equation}
\begin{aligned}
  D_i(t) &=2\Re\left[d_i(t)e^{-i\omega_it}\right] \\ 
  &=2\Re\left[ d_i^{(0)}(t)e^{-i\omega_it} + \sum_j\frac{\Delta_{ij}}{2J_{ij}^*}g_{ij}(t)e^{-i\omega_jt}\right],
  \label{eq:fixed_drive}
  \end{aligned}
\end{equation}
where $d_i^{(0)}(t), g_{ij}(t)$ are slow functions (compared to the coupling rate), we have defined the qubit--qubit detuning $\Delta_{ij}=\omega_i-\omega_j$ and the sum over $j$ is taken to run over coupling qubits for which $J_{ij}\neq0$.

We can now separate the Hamiltonian into a slow and a fast part. 
The slow Hamiltonian corresponds to the single-qubit Hamiltonian derived in the main text~\cref{eq:h(t)_lab}, except for many qubits
\begin{equation}
  H_{\mathrm{slow}} = \sum_i d_i^{(0)}(t)\sigma_+^{(i)}+\mathrm{H.c.}
  \label{eq:fixed_slow}
\end{equation}
The fast part is off-resonant and has little effect, except when there are resonant processes.
For example, to second order, the hopping becomes resonant again, leading to an Ising term of order $|J_{ij}|^2/\Delta_{ij}$.
Since our original assumption is that the coupling is weak compared to the detuning, $J_{ij}\ll\Delta_{ij}$, this is negligible.
However, the other term that appears at second order, mixing the off-resonant drive $g_{ij}$ with the hopping $J_{ij}$, is not negligible and in fact gives rise to the interaction~\cite{Paraoanu2006,Rigetti2010,Krantz2019}
\begin{equation}
  H_{\mathrm{int}}^{(2)} = \sum_{\langle ij\rangle }g_{ij}(t)\sigma_z^{(i)}\sigma_+^{(j)}+\mathrm{H.c.}
  \label{eq:fixed_int}
\end{equation}
Combining \cref{eq:fixed_slow,eq:fixed_int} allows one to engineer a wide range of time-dependent Hamiltonians.
Note that due to drive enabling a resonant interaction between coupled qubits, the disorder has disappeared. 
The disorder can be restored by detuning the ``coupling drive'' $g_{ij}(t) = \exp(i\int^th_z^{(j)}(t')dt')\tilde g_{ij}(t)$ and performing the same transformation on the single-qubit drives $d_i^{(0)}(t) = \exp(i\int^th_z^{(i)}(t')dt')\tilde d_i^{(0)}(t)$,
which in a rotating frame with respect to 
$H_z = \sum_i h_z^{(i)}\sigma_z^{(i)}$ reads
\begin{equation}
  H = \sum_i \vec h_i(t)\cdot\vec\sigma_i + \sum_{\langle ij\rangle }\tilde g_{ij}^x(t)\sigma_z^{(i)}\sigma_x^{(j)}
  +\tilde g_{ij}^y(t)\sigma_z^{(i)}\sigma_y^{(j)}
  \label{eq:fixed_final_H}
\end{equation}
with
\begin{equation}
  \tilde g_{ij}(t) =\frac12\left[ \tilde g_{ij}^x(t)-i\tilde g_{ij}^y(t)\right],
  \label{eq:tilde_g}
\end{equation}

The induced coupling is rich, as we have independently controllable $XZ, ZX, YZ$, and $ZY$ interactions.
A special case arises if the qubit coupling graph is bipartite, i.e., if it can be separated into two subgraphs $A$ and $B$ such that nodes from $A$ are only connected to $B$ and vice versa, as for example in a square lattice.
In this case, we can make a unitary transformation on every qubit in one of the sublattices ($B$, say) that rotates $Z$ into $X$ and $X$ into $-Z$. 
This allows us to engineer two particularly important interactions
\begin{itemize}
  \item[(i)] Ising: Starting from only $XZ$ or $ZX$ only, the unitary transformation yields $XX$ or $ZZ$.
  \item[(ii)] XY model: Applying both $XZ$ and $ZX$, the unitary transformation yields $g_1XX+g_2ZZ$, or equivalently $g_1XX+g_2YY$, where the coupling strengths are fully tuneable.
\end{itemize}

Finally, we note that the biggest limitation for the controllability of this setup an effect similar to \emph{frequency crowding}, leading to crosstalk.
As this approach requires one to apply a range of different control tones at different frequencies, unwanted resonances can occur, which may affect the final effective Hamiltonian for the system.
This problem particularly concerns lattices with high connectivity, but can typically be avoided by careful consideration and design of the applied tones~\cite{Rigetti2010}.

\subsection{Tuneable frequencies, but fixed interaction}

If the qubit frequencies are tuneable, all the techniques from the previous section can be employed, but one gains the additional capacity to preclude frequency crowding.

In addition, one can make use of the direct interaction of the qubits by bringing them close to resonance. This realizes an XY interaction with tuneable disorder.
\begin{equation}
  H = \sum_i h_z^{(i)}\sigma_z + \sum_{\langle ij\rangle }J_{ij}(\sigma_+^{(i)}\sigma_-^{(j)}+\mathrm{H.c.})
  \label{eq:tuneable_frequencies}
\end{equation}

\subsection{Tuneable interactions}

For a set of qubits that interact via a tuneable XY interaction, the full Hamiltonian reads
\begin{equation}
  \begin{aligned}
  	H(t) &= \sum_i\omega_ia_i\dagg a_i+U_ia_i\dagg a_i\dagg a_ia_i+(a_i+a_i\dagg)D_i(t)\\
  	&+\sum_{\langle ij\rangle }g_{ij}(t)(a_i\dagg a_j+a_j\dagg a_i),
  \end{aligned}
  \label{eq:full_array}
\end{equation}
Parameterizing the drive as before in terms of an effective magnetic field $\vec h_i(t)$, passing to a rotating frame with respect to the Hamiltonian $H_0=\sum_i(\omega_0-h^{(i)}_z(t))a_i\dagg a_i$ and restricting to the qubit subspace, we find
\begin{equation}
  \begin{aligned}
  	&H(t) = \sum_i (\omega_i-\omega_0+h^{(i)}_z(t))\sigma_z^{(i)}
  	+\left\{\frac{1}{2}\sigma_-^{(i)}h_+^{(i)}(t)\right.\\
  	&\left.+\sum_{\langle ij\rangle }g_{ij}(t)\sigma_+^{(i)}\sigma_-^{(j)}e^{2i\int_0^t [h_z^{(j)}(t')-h_z^{(i)}(t')]dt'}+\mathrm{H.c.}\right\}
  \end{aligned}
  \label{eq:array_rotating}
\end{equation}
Choosing $h_z^{(i)}(t)=h_z(t)$ for all $i$ removes the phase on the interaction terms. 
The resulting Hamiltonian is
\begin{equation}
  \begin{aligned}
  H(t) &= \sum_i h_x^{(i)}(t)\sigma_x^{(i)}+h_y^{(i)}(t)\sigma_y^{(i)}+
  [h_z(t)+\delta_i]\sigma_z^{(i)}\\
  &+\sum_{\langle ij\rangle }g_{ij}(t)\left[ \sigma_x^{(i)}\sigma_x^{(j)}+\sigma_y^{(i)}\sigma_y^{(j)} \right].
  \end{aligned}
  \label{eq:full_result}
\end{equation}
This is the Hamiltonian quoted in the main text.
The disorder in the energies is inherited from the intrinsic disorder of the platform.

\section{Brief review of the half-BHZ model}

The half-Bernevig-Hughes-Zhang (BHZ) is one of the simplest models for a two-dimensional Chern insulator~\cite{BernevigBook2013}. It is a two-band lattice model that is most conveniently described in momentum space by the Hamiltonian
\begin{equation}
\hat{H}(\vec{k}) = \sin^{} (k_x)\hat{\sigma}_x + \sin^{} (k_y) \hat{\sigma}_y + B\left[M - \cos^{} (k_x) - \cos^{} (k_y) \right] \hat{\sigma}_z.
\end{equation}
This Hamiltonian has the form $\hat{H}(\vec{k}) = \vec{h}(\vec{k})\cdot \vec{\sigma} = \sum_a h_a(\vec{k}) \hat{\sigma}_a$, and spectrum given by $\lambda_{\pm} = \pm |\vec{h}|$, or more explicitly,
\begin{equation}
    \lambda_{\pm} = \pm \sqrt{\sin^2(k_x) + \sin^2(k_y) + B^2\left[M - \cos^{}(k_x) - \cos^{}(k_y) \right]^2 }
\end{equation}
which is gapped except at $M=-2,0,2$. Depending on the value of $M$ this model realizes three different phases with these gap closings marking the phase transitions between them. For $|M|>2$ we have a trivial insulator. For $-2 < M < 0$, we have Chern insulator with Chern number $C=+1$, and $0 < M < 2$ describes a Chern insulator with $C=-1$.

The eigenstates for the upper and lower band have the particularly simple form
\begin{equation}
    |\psi_{\pm}\rangle = \frac{1}{\sqrt{2h(h\pm h_z)}}
\left(\begin{array}{c}
    h \pm h_z \\
    h_x + i h_y
    \end{array}\right),
\end{equation}
which correspond to vectors on the Bloch sphere (anti-)aligned with the field direction $\vec{h}$. This can easily be seen by considering the vector of expectation values $\langle \vec{\sigma} \rangle = \pm \vec{h}/h$.

\subsection{Chern number}

\begin{figure}[t]
    \centering
    \includegraphics[width=\columnwidth]{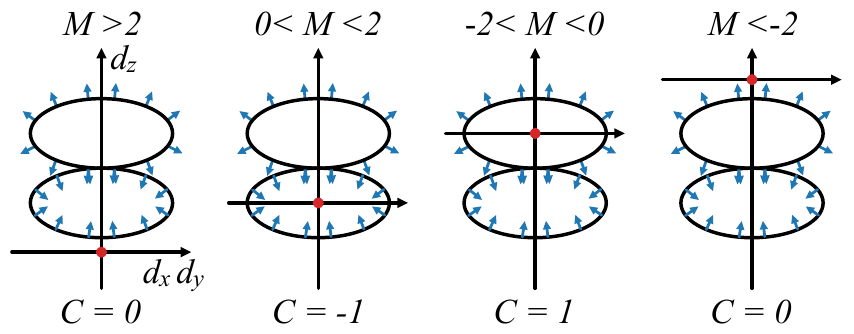}
    \caption{Schematic of surface generated by the map $\vec{h}(\vec{k})$ from the Brillouin zone. A two-dimensional cross-section in corresponding to constant $k_x$ or $k_y$ is shown for simplicity. The normal to the surface is indicated by the blue arrows and the origin is highlighted in red.}
    \label{fig: chern schematic}
\end{figure}

To compute the Chern number we require the Berry connection $A_i(\vec{k})$ and/or the Berry curvature $F_{ij}(\vec{k})$, defined by 
\begin{equation}
\begin{aligned}
    A_i(\vec{k}) &= \sum_{\psi \in \text{filled bands}} -i\langle \psi | \frac{\partial}{\partial k_i} |\psi \rangle, \\
    F_{ij}(\vec{k}) &= \frac{\partial A_j}{\partial k_i} - \frac{\partial A_i}{\partial k_j}.
\end{aligned}
\end{equation}
The Chern number $C \in \mathbb{Z}$ is then computed by integrating the Berry curvature over the Brillouin zone:
\begin{equation}
    C = \frac{1}{2\pi} \iint_\text{B.Z.} \text{d}k_x \text{d}k_y F_{xy}(\vec{k}).
\end{equation}
For the two-band model, the Berry curvature has a particularly intuitive form, namely
\begin{equation}
    F_{xy}(\vec{k}) = \frac{1}{2h^3} \epsilon^{abc} h_a \partial_{k_x} h_b \partial_{k_y} h_c = \frac{1}{2h^3} \vec{h} \cdot (\partial_{k_x} \vec{h} \times \partial_{k_y} \vec{h} ).
\end{equation}
The vector $\vec{h}$ defines a map from the torus (B.Z.) to a two-dimensional surface $M$ embedded in three dimensions. The Berry curvature then corresponds to the dot product of a field due to a unit charge at the origin and the normal of the surface $S$. The Chern number therefore measures the flux through the surface $S$, that is
\begin{equation}
    C = \frac{1}{4\pi} \iint_M \frac{\vec{h}}{h^3} \cdot \text{d}\hat{S},
\end{equation}
which measures the degree of map defined by $\vec{h}$, or equivalently, the winding of the surface $M$ around the origin.

The surface prescribed by $\vec{h}(\vec{k})$ is closed self-intersecting surface with two lobes. the upper (along $z$-axis) lobe has normal facing outward, while lower has inward facing normal. The parameter $M$ controls the position of this surface along the $z$-axis. The three phases therefore correspond to when the origin falls outside of the surface, or within one of the two lobes. This is shown schematically using a two-dimensional cross section in Fig.~\ref{fig: chern schematic}. Given this interpretation we can see that when $|M|>2$ the Chern number must be zero since $h_z$ does not change sign.

\section{Floquet lattice}\label{app:floquet_lattice}

The connection to the BHZ Chern insulator is elucidated by considering the Floquet construction which relates our time-dependent zero-dimensional system with a two-dimensional lattice Hamiltonian. This is done by considering the following the time-dependent states
\begin{equation}\label{eq: floquet psi}
    |\Psi(t) \rangle = e^{-iEt} \sum_{\alpha,\vec{n}} \psi^\alpha_{\vec{n}} e^{-i\vec{n}\cdot\vec{\omega} t} |\alpha \rangle,
\end{equation}
where $\alpha=0,1$ label the two basis states for the qubit, and $\vec{n} = (n_1, n_2)$, with $n_i \in \mathbb{Z}$ and $\vec{\omega} = (\omega_1,\omega_2)$. The vector $\vec{n}$ is associated with the sites on an infinite two-dimensional square lattice. The time-dependent Schr{\"o}dinger equation $i\partial_t |\Psi(t)\rangle = H(t) |\Psi(t)\rangle$, can then be written as
\begin{equation}
    \begin{aligned}
    	E \psi_{\vec{k}} = \Big\{&\sin^{}(k_x + \phi_1) \sigma_x + \sin^{}(k_y+\phi_2) \sigma_y\\
    	&+ \left[M-\cos^{}(k_x+\phi_1) - \cos^{}(k_y+\phi_2)\right]\sigma_z \Big\} \psi_{\vec{k}}\\ 
    &- \sum_{\vec{n}}e^{-i\vec{k}\cdot \vec{n}} \vec{n}\cdot\vec{\omega} \, \psi_{\vec{n}},
    \end{aligned}
\end{equation}
where $\psi_{\vec{k}} = \sum_{\vec{n}} e^{-i \vec{k} \cdot \vec{n}} \psi_{\vec{n}}$, and $\vec{k} = (k_1, k_2)$ with $k_i \in [0,2\pi)$.
The first three terms correspond to the half-BHZ Chern insulator. The final term is in mixed form and corresponds to a linear potential along the direction $\vec{\omega}$, which can be interpreted as an electric field. In the strong-drive limit, $\eta \gg \omega_1,\omega_2$, this induces a time dependence in the momentum $\vec{k} \rightarrow \vec{k}_0 + \vec{\omega} t$~\cite{Martin2017}. Making the substitution for the time-dependent momentum we recover Eq.~\eqref{eq:h(t)}. The dynamics corresponds to the adiabatic evolution of the ground state of Eq.~\eqref{eq:h(t)} in the Brillouin zone, so long as the gap controlled by $M$ is sufficiently large compared with $\omega_1$.

\begin{figure}[tb]
  \includegraphics[width=\columnwidth]{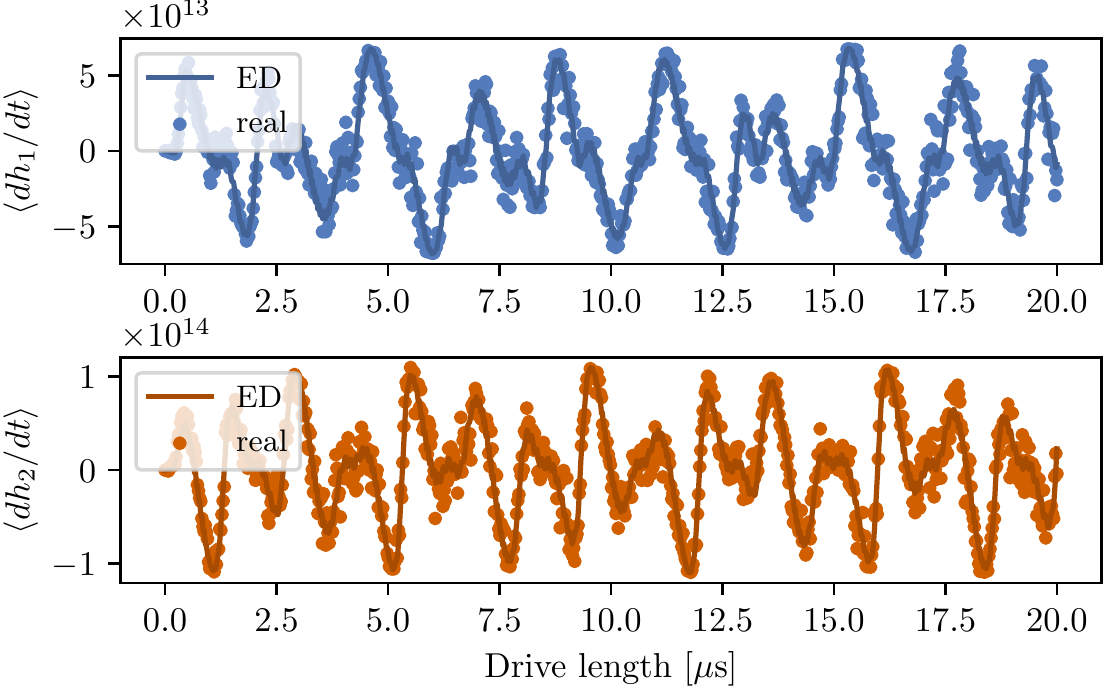}
\caption{Expectation values of $\langle \text{d} h_i / \text{d} t \rangle$ computed from the projected tomographic data shown in Fig.~\ref{fig:tomography}. The experimental data (circular markers) are compared with exact numerical simulations (solid lines).  }\label{fig: h dot}
\end{figure}

\section{Measuring the work done by each drive}\label{app:measuring}

The computation of the work done by each drive proceeds by three main steps: i) state tomography, to obtain the spin expectation values $\langle \vec{\sigma} \rangle$ at each time; ii) compute the expectation value $\langle \text{d} h_i / \text{d} t \rangle$; iii) numerically integrate Eq.~\eqref{eq:W_i} to get the work done by each drive.

To perform state tomography we run the experiment three times (with 8192 shots each) and measure the qubit in the $X$, $Y$ and $Z$ basis. We take the computational or natural basis for the qubit to be the $Z$ basis. Each measurement gives an outcome of 0 or 1 corresponding to the eigenstates of the Pauli-$Z$ operator with positive $(+1)$ and negative $(-1)$ eigenvalues, respectively. By averaging over the 8192 shots we obtain the average expectation value $\langle Z \rangle$, with standard error of the mean $1/\sqrt{8192} \sim 1\%$. To measure in the $X$ basis we apply a Hadamard gate before measuring the qubit, and for the $Y$ basis we apply an $S^\dag$ gate and a Hadamard gate before measuring. We then move to the rotating frame using Eq.~\eqref{eq: z rotation}. Fig.~\ref{fig:tomography} from the main text shows an example of the measured Pauli expectation values corresponding to the data shown in Fig.~\ref{fig:frequency_conversion} (i.e., $M=1$) compared with the exact numerical simulation. The measured expectation values will not in general be consistent with a pure state due to a range of factors including statistical errors, measurement errors, and decoherence. We therefore normalize the vector of expectation values $\langle \vec{\sigma} \rangle$, which corresponds to projecting onto the Bloch sphere. 

From this projected tomography data we can then compute the expectation values $\langle \text{d} h_i / \text{d} t \rangle =  \text{d} \vec{h}_i / \text{d} t \cdot \langle \vec{\sigma} \rangle$, where 
\begin{equation}
\begin{aligned}
    \frac{\text{d} \vec{h}_1}{\text{d} t} = \eta \omega_1 \cos (\omega_1 t + \phi_1) \hat{x} + \eta \omega_1 \sin (\omega_1 t + \phi_1) \hat{z}, \\
    \frac{\text{d} \vec{h}_2}{\text{d} t} = \eta \omega_2 \cos (\omega_2 t + \phi_2) \hat{y} + \eta \omega_2 \sin (\omega_2 t + \phi_2) \hat{z}, 
\end{aligned}
\end{equation}
where, e.g. $\hat{x} \cdot \langle \vec{\sigma}\rangle = \langle \sigma_x \rangle$. In Fig.~\ref{fig: h dot} we show the expectation values computed from the Pauli expectation values shown in Fig.~\ref{fig:tomography}, corresponding to Fig.~\ref{fig:frequency_conversion} of the main text.

The final step is to numerically integrate these expectation values to get the work done by each drive, as given in Eq.~\eqref{eq:W_i}. We use a simple trapezoid rule using data sampled at 800 points in time to 20 $\mu$s. 

In Fig.~\ref{fig: all work done} we show the data from which we extracted the work done by the drives to produce Fig.~\ref{fig:chern_number}.

\section{Simulation details}

\begin{figure}[t!]
  \includegraphics[width=\columnwidth]{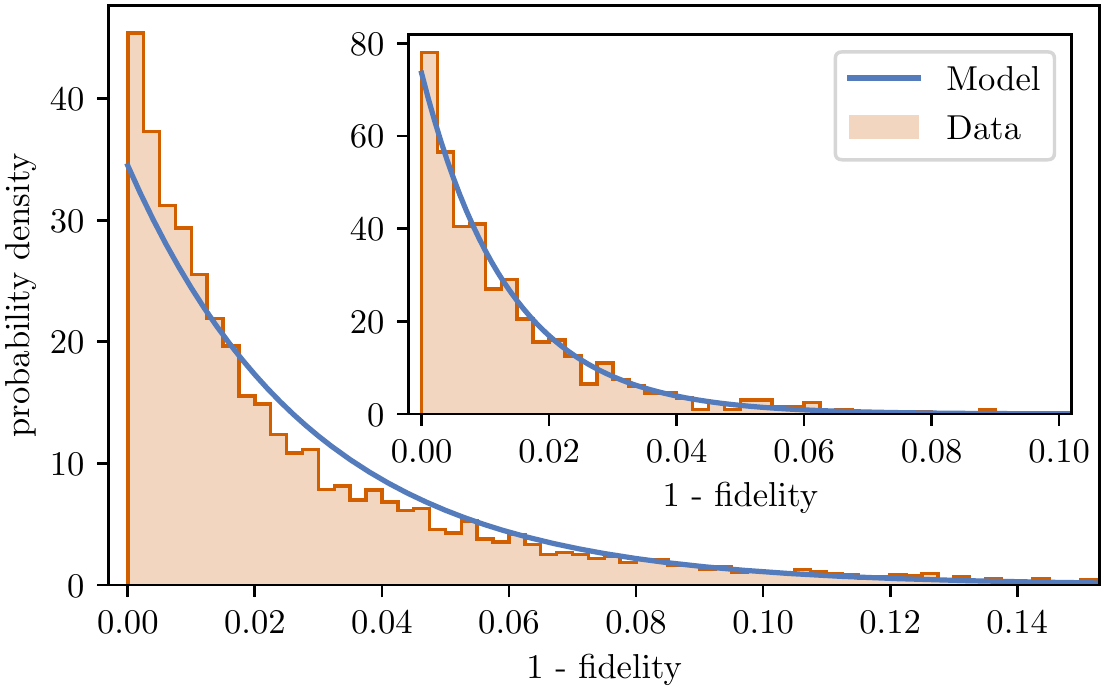}
\caption{Distribution of the state fidelity measured across all experiments, i.e., across all drive lengths and values of $M$ (excluding $M=1.7,2.3$ due to diabatic effects). The distribution is compared with an exponential distribution set by the average fidelity, which is used for our heuristic error model. Inset is the corresponding distribution for a specific value of $M=1$.}\label{fig: fidelity distribution}
\end{figure}

\subsection{Exact numerical simulation}

Th IBM device can implement a time-dependent drive of the form Eq.~\eqref{eq:drive_parameterization} given in the main text, where the drive shape $d(t)$ is piece-wise constant with interval $dt=0.22$ns. For comparison with experimental results we simulate the corresponding effective Hamiltonian $H(t) = \vec{h}(t) \cdot \vec{\sigma}$ with piece-wise constant $\vec{h}(t)$ using exact numerical simulations. This numerical simulation consists of discretizing time with interval $dt=0.22$ns and performing time-independent unitary evolution during each period using exact diagonalization. 

\subsection{Heuristic error model for fidelity}

In Fig.~\ref{fig:frequency_conversion} of the main text we include a shaded region that corresponds to a simple error model based on the measured fidelity of the state of the qubit. Here we give motivation for and details of this simple error model.

We define the fidelity as 
\begin{equation}\label{eq: fidelity SI}
    \mathcal{F} = |\langle \psi(t) | \phi (t) \rangle|^2,
\end{equation}
where $|\phi(t)\rangle$ is the state at time $t$ according to our exact numerical simulation, and $|\psi(t)\rangle$ is the measured state of the qubit at time $t$ after projecting back onto the Bloch sphere. We find that the fidelity across the entire length of the drive is approximately distributed according to an exponential distribution, i.e. $p(x) = \frac{1}{\beta} e^{-x/\beta}$, where $\mathbb{E}[p(x)] = \beta$. 
In Fig.~\ref{fig: fidelity distribution} we show the distribution of the fidelity across all experiments (neglecting $M=1.7,2.3$), as well as for a given value of $M=1$ (inset). For both a given value of $M$ and across all experiments, the exponential distribution matches well the experimentally measured distribution of fidelities. Note that the averaged distribution is slightly less well captured since the average of exponential distributions with different means is not an exponential distribution. We additionally find that the distribution of fidelities is approximately independent of drive length over the time scales we consider, as shown in Fig.~\ref{fig: fidelity}, suggesting that measurement and statistical errors are dominant. Note that we neglect $M=1.7,2.3$ due to strong diabatic effects, see below.

Based on the average measured fidelity of 0.971 we use a simple heuristic error model. This involves perturbing the exact numerical simulations (sampled at 800 points in the 20$\mu$s interval) such that we produce an exponential distribution of the fidelity. To achieve the correct distribution we first note that
\begin{equation}
|\langle \psi | \phi \rangle|^2 = \frac{1 + \vec{\psi}\cdot \vec{\phi} }{2} = \frac{1 + \cos \theta_{\psi,\phi}}{2},
\end{equation}
where $\vec{\psi} = (\langle \sigma^x \rangle_\psi, \langle \sigma^y \rangle_\psi, \langle \sigma^z \rangle_\psi)^T$, is the corresponding 3D unit Bloch vector. For the fidelities to be correctly distributed, we sample $\theta_{\psi,\phi}$ according to
\begin{equation}\label{eq: theta}
    \theta_{\psi,\phi} = \arccos (1-2\min(x,1)),
\end{equation}
where $x$ is randomly sampled from $p(x) = \frac{1}{\beta} e^{-x/\beta}$, with $\beta=0.029$. We include the $\min$ function in Eq.~\eqref{eq: theta} to exclude rare events in the tail of the exponential distribution that would correspond to $\mathcal{F} <0$. Given the exact state $|\phi\rangle$ we generate $|\psi\rangle$ by
\begin{equation}
    |\psi\rangle = R_{\vec{n}} (\theta_{\psi,\phi}) |\phi \rangle,
\end{equation}
where $\vec{n}$ is a random vector uniformly distributed on the circle orthogonal to $\vec{\phi}$, and $R_{\vec{n}} (\theta)$ is a single qubit rotation by $\theta$ around $\vec{n}$.

In practice we work with the 3D Bloch vectors $\vec{\psi}$. The rotation matrix is then given by
\begin{equation}
    R_{\vec{n}}(\theta) = 1 + \sin\theta \, A + (1-\cos\theta) \, A^2,
\end{equation}
where
\begin{equation}
    A = \left( \begin{array}{ccc}
    0 & -n_z & n_y \\
    n_z & 0 & -n_x \\
    -n_y & n_x & 0 
    \end{array}\right).
\end{equation}
We generate $\vec{n}$ by sampling each element $n_i$ from a normal distribution with zero mean and unit variance, then normalize to get a vector uniformly sampled from the unit sphere. We then orthogonalize this vector to $\vec{\phi}$ and normalize again to get a vector uniformly sampled from the circle orthogonal to $\phi$.

To produce the shaded region in Fig.~\ref{fig:frequency_conversion} we use our simple error model to produce "fake data". We then treat this in the same way we treat the experimental data to extract the Chern number. Using 500 random realizations we obtain a distribution for the extracted values of the Chern number. The shaded region then corresponds to $\bar{C} \pm \sigma$, where $\bar{C}$ is the mean and $\sigma$ is the sampled standard deviation. Importantly, we find that very small average error rate $1-\mathcal{F} \approx 3\%$ translates into large standard deviation of approximately 0.25 for the extracted Chern number.

\subsection{Underlying error model}

The observed heuristic error model for the fidelity can be accounted for by a simple model of noisy measurements in the device. This noise could be due to a number of sources including noise during the driving or measurement process, or due to the random drift of systematic readout errors between experiments. 

Let us assume that the measurement process is subject to Gaussian noise such that
\begin{equation}
    \langle \vec{\sigma} \rangle_{\text{real}} = \langle \vec{\sigma}\rangle_{\text{exact}} + \vec{\eta},
\end{equation}
where $\eta \sim \mathcal{N}(\mu=0,\sigma_{\mathrm{noise}})$ follows a normal distribution with zero mean $\mu$ and standard deviation $\sigma_{\mathrm{noise}}$. Note that this model can result in impossible values for the expectation values, but for our analysis we focus on typical states where $\langle \sigma_i\rangle_{\text{exact}} \neq 0, 1$, and $\sigma_{\mathrm{noise}} \ll 1$.

Let us now look at what distribution for the fidelity this model results in. Let us define $\vec{u} = \langle \vec{\sigma} \rangle_{\text{exact}}$, which is a 3D unit vector containing the exact expectation values, and $\vec{v} = \langle \vec{\sigma} \rangle_{\text{real}} / |\langle \vec{\sigma} \rangle_{\text{real}}|$, is the vector measured after our noise model project back to the Bloch sphere. The fidelity is then given by
\begin{equation}
    \mathcal{F} = \frac{1 + \vec{u}\cdot \vec{v}}{2} = \frac{1 + \cos\theta}{2}.
\end{equation}
For $\sigma_{\mathrm{noise}}\ll 1$, we can approximate $\theta$ by the orthogonal component of the perturbation generated by the noise, that is $\theta \approx |\eta_\perp|$. Therefore we have that
\begin{equation}
    1 - \mathcal{F} \approx \frac{|\eta_\perp|^2}{4}.
\end{equation}
Since $\eta_\perp$ is sampled from a 2D Gaussian distribution, its norm squared, $|\eta_\perp|^2$, is distributed according to a chi-squared distribution for two degrees of freedom, i.e., $\chi^2(k=2)$. This distribution is given by an exponential distribution with PDF $p(x)=\frac{1}{\lambda}e^{-x/\lambda}$, where $\lambda = \frac{1}{2} \sigma_{\mathrm{noise}}^2$, which reproduces the distribution of fidelity loss measured experimentally. Notably, the average value of the fidelity corresponds to $\sigma_{\mathrm{noise}} \approx 0.24$, indicating a large amount of noise in the device ($12\%$ error). Note that the statistical errors for 8192 shots would correspond to a standard deviation an order of magnitude smaller than this, and so we can consider statistical errors negligible.

\section{Experimental data}

Here we plot extra data to supplement what is shown in the main text. Note that all data is hosted at Zenodo and freely available~\cite{Note1}.

\subsection{Purity}

\definecolor{LightRed}{rgb}{1,0.92,0.9}
\begin{table}[bh!]
\centering
\begin{tabularx}{\columnwidth}{| >{\centering\arraybackslash}X | >{\centering\arraybackslash}X | >{\centering\arraybackslash}X | >{\centering\arraybackslash}X |}
    \hline
    $M$ & $\overline{\text{purity}}$ & $\tau$ & $\overline{\text{fidelity}}$  \\
    \hline
    0.6 & 0.850 & 75.5$\mu$s & 0.963 \\
    0.8 & 0.873 & 107$\mu$s & 0.983 \\
    1 & 0.855 & 109$\mu$s & 0.986 \\
    1.2 & 0.879 & 195$\mu$s & 0.987 \\
    1.4 & 0.872 & 265$\mu$s & 0.986 \\
    \rowcolor{LightRed}
    1.7 & 0.829 & 29.3$\mu$s & 0.836 \\
    \rowcolor{LightRed}
    2.3 & 0.806 & 24.9$\mu$s & 0.752 \\
    2.6 & 0.875 & 97.5$\mu$s & 0.950 \\
    2.8 & 0.889 & 977$\mu$s & 0.973 \\
    3 & 0.873 & 158$\mu$s & 0.965 \\
    3.2 & 0.866 & 89.7$\mu$s & 0.951 \\
    3.4 & 0.894 & 272$\mu$s & 0.970 \\
    \hline
    average & 0.863 & 200$\mu$s & 0.942 (0.971) \\
    \hline
\end{tabularx}
\caption{Average purity (see Fig.~\ref{fig: purity}), decoherence time, and fidelity (see Fig.~\ref{fig: fidelity}) of our experimental data. The decoherence time is extracted from an exponential fit to the purity data. While this extracted time scale is not an accurate reflection of the decoherence time of the devices due to the large purity fluctuations observed, the average value of $200\mu$s matches the values $T1,T2 \gtrsim 100 \mu$s measured by IBM. The highlighted rows correspond to the data that differs significantly from the numerical simulation (see Fig.~\ref{fig: fidelity}). The fidelity in brackets corresponds to the average when disregarding data for $M=1.7$ and $2.3$, which appear to be strongly affected by diabatic effects (see Fig.~\ref{fig: fidelity}).}\label{tab: errors}
\end{table}

\begin{figure}[t!]
  \includegraphics[width=\columnwidth]{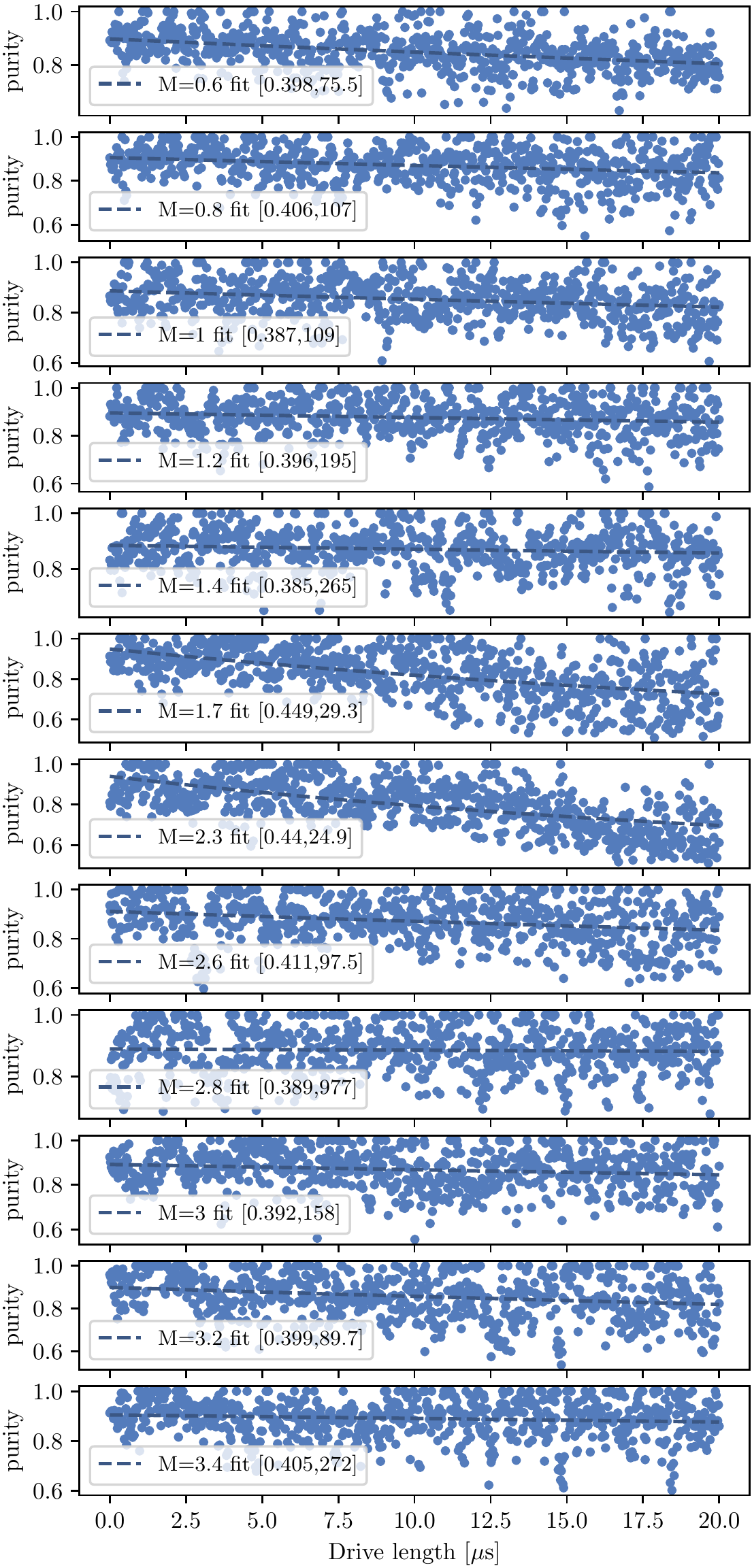}
\caption{Purity $\text{Tr}(\rho^2)$ of the state constructed from the experimentally measured Pauli expectation values. We fit the purity data with functional form $1/2 + ae^{-t/\lambda}$, with the fit parameters $[a,\lambda]$ shown in the corresponding subfigures, where $\lambda$ is given in $\mu$s.}\label{fig: purity}
\end{figure}

\begin{figure}[t!]
  \includegraphics[width=\columnwidth]{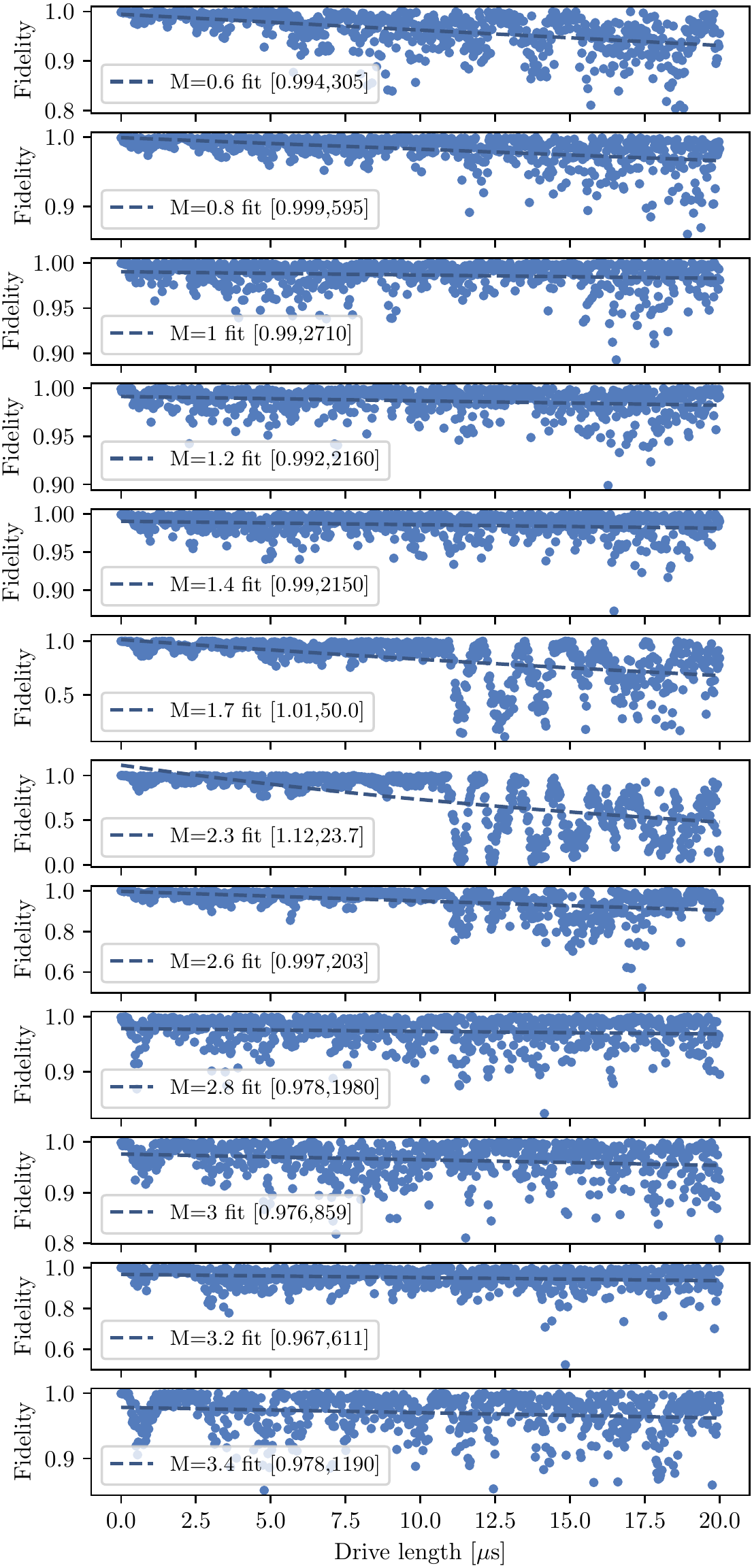}
\caption{Fidelity of the measured state after projecting onto the Bloch sphere compared with the exact numerical simulation. Explicitly, we project the density matrix $\rho(t)$ extracted from the Pauli expectation values at time $t$ onto the Bloch sphere to obtain the pure state $|\psi(t)\rangle$. The fidelity is then $\mathcal{F} = |\langle\psi(t) | \phi(t)\rangle|^2$, where $|\phi(t)\rangle$ is the exact numerical simulation of the state at time $t$. We fit the fidelity with functional form $be^{-t/\xi}$, with the fit parameters $[b,\xi]$ shown in the corresponding subfigure, where $\xi$ is given in $\mu$s. For $M=1.7,2.3$ we observe a sudden dramatic drop in the fidelity after $10\mu$s, which is reflected in the behaviour of the work done shown in Fig.~\ref{fig: all work done}. }\label{fig: fidelity}
\end{figure}

\begin{figure*}[t!]
  \includegraphics[width=\textwidth]{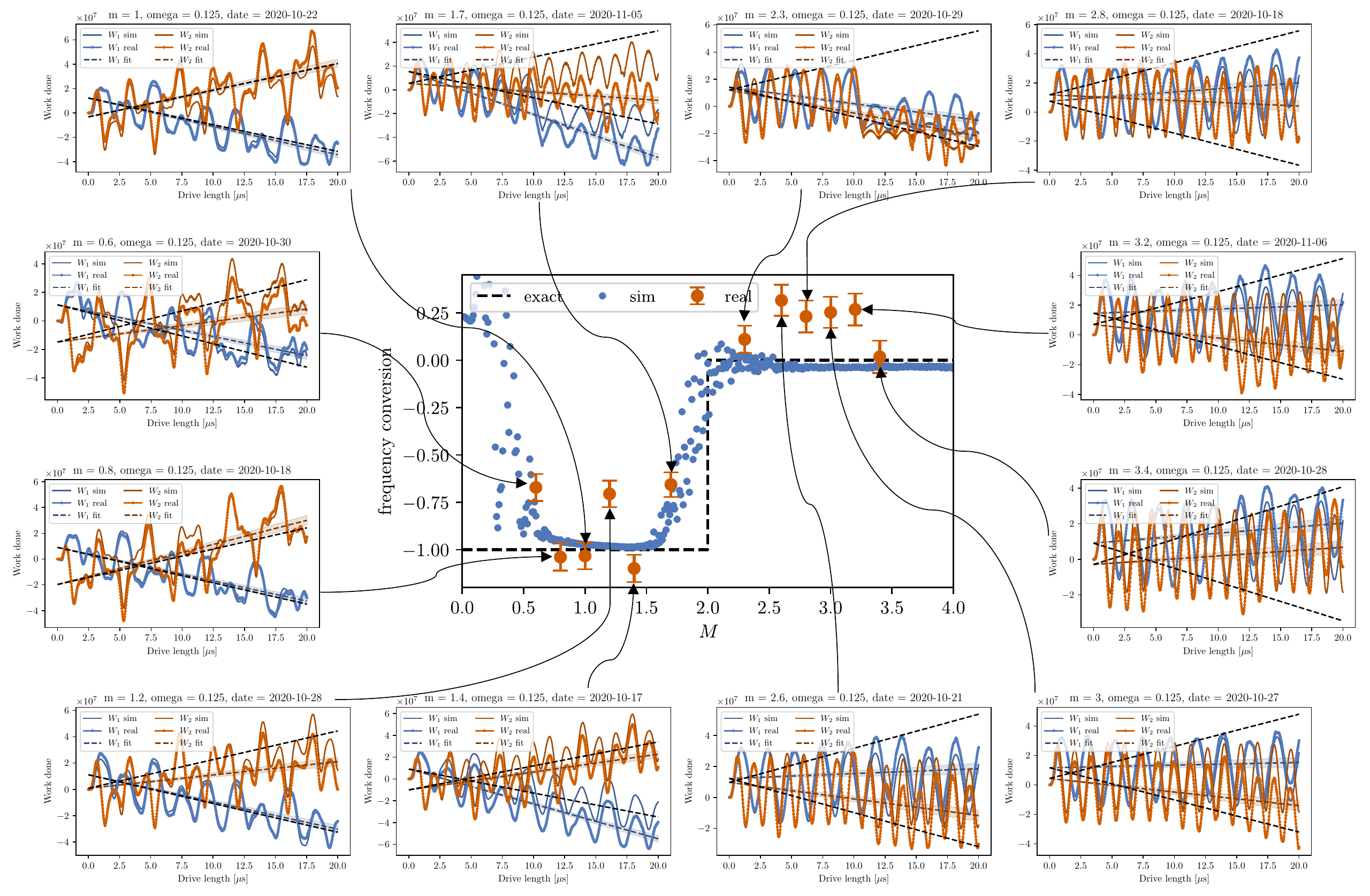}
\caption{Plots of the work done by the two drives and their linear fits for each of the data points shown in Fig.~\ref{fig:chern_number}, as discussed in \cref{app:measuring} and~\cref{ap: work done}}\label{fig: all work done}
\end{figure*}

To compute the purity of the measured state, we first construct the density matrix from the measured Pauli expectation values as follows
\begin{equation}
    \rho = \frac{1}{2} \left(1 + \langle \sigma^x \rangle \sigma^x + \langle \sigma^y \rangle \sigma^y + \langle \sigma^z \rangle \sigma^z \right).
\end{equation}
However, this density matrix is not necessarily physical, and can correspond to a Bloch vector with greater than unit length. To account for this we find the closest physical density matrix using maximum likelihood~\cite{Smolin2012}. For a single qubit, this is equivalent to projecting any unphysical state onto the Bloch sphere. From the physical density matrix we can then compute the purity
\begin{equation}
    \text{tr}(\rho^2) = \frac{1}{2} + \frac{1}{2} \left(\widetilde{\langle \sigma^x \rangle}^2 + \widetilde{\langle \sigma^y \rangle}^2 + \widetilde{\langle \sigma^z \rangle}^2 \right),
\end{equation}
where $\widetilde{\langle\sigma^i\rangle}$ is the corrected expectation value.

In Fig.~\ref{fig: purity} we show the measured purity of the qubit for the different values of $M$. We fit this data with an exponential form $1/2 + a e^{-t/\lambda}$. Additionally, we present the average purities and the time scale $\lambda$ extracted from the fit in Table~\ref{tab: errors}. Across all experiments we find an average purity of $0.863$. We also find that the extracted decoherence time $\lambda \approx 200 \mu$s, closely matches the values for the device measured independently by IBM, namely, $T_1, T_2 \gtrsim 100 \mu$s. However, we note that due to the large fluctuations of the measured purity this is not a reliable measure of the decoherence time of the device. Furthermore, the decoherence times quoted by IBM change on a daily basis, and in particular between runs for different $M$.

\subsection{Fidelity}

In Fig.~\ref{fig: fidelity} we show the fidelity [see \cref{eq: fidelity SI}] of the error mitigated states at each time and for each value of $M$. 

For the values of $M=1.7,2.3$ we observe large oscillations in the fidelity not seen for other values of $M$. This suggests that for these values there are strong diabatic effects. This behaviour is also reflected by a sudden change in work done that is also visible in the numerical simulation, see Fig.~\ref{fig: all work done}. This is consistent with diabatic effects due to the small gap close to $M=2$, and was also seen in the numerical simulations by Martin {\it et al}~\cite{Martin2017}. Small changes in the initial state along with inexact driving in the experiment lead to the fidelity oscillations observed. For this reason we exclude the these two values of $M$ when discussing the average fidelity across our experiments.

For all other values of $M$ we achieve an average fidelity of 0.971.
Note that, as explained above, this corresponds to a 12\% error in the expectation values.
Furthermore, we observe that the fidelity remains approximately constant over this time scale, with an extracted decay time approximately an order of magnitude greater than the $T_1$ and $T_2$ times of the device. This supports our use of error mitigation by projecting on to the Bloch sphere. Note, that this slow decay and the effectiveness of the error mitigation can only be maintained while our total drive length is significant less than the decoherence time of the device. It also motivates our use of a time-independent distribution for the fidelity in our heuristic error model described in the previous section.

\subsection{Work done}\label{ap: work done}

In Fig.~\ref{fig: all work done} we include plots for the work done for each of the values of $M$. From the work done we extract the Chern number as plotted in the centre of the figure, which is a reproduction of Fig.~\ref{fig:chern_number} from the main text. The Chern number is extracted using a linear fit for the work done by the two drives, according to Eq.~\eqref{eq:work quantization} of the main text.

\bibliography{library,editable_references.bib}
\end{document}